\documentclass[aps,prd,final, nofootinbib]{revtex4}

\usepackage{amsmath, amsfonts, amsthm, amssymb, graphicx,epsfig}
\usepackage{subfig}
\usepackage{epstopdf}
\usepackage{color}
\def\be{\begin{equation}}
\def\ee{\end{equation}}
\def\ba{\begin{eqnarray}}
\def\ea{\end{eqnarray}}

\def\bdm{\begin{displaymath}}
\def\edm{\end{displaymath}}

\def\bq{\begin{quote}}
\def\eq{\end{quote}}

\def\d{{\rm d}}
\def\del{\partial}
\def\ltap{\ \raise.3ex\hbox{$<$\kern-.75em\lower1ex\hbox{$\sim$}}\ }
\def\gtap{\ \raise.3ex\hbox{$>$\kern-.75em\lower1ex\hbox{$\sim$}}\ }
\def\gl{\ \raise.5ex\hbox{$>$}\kern-.8em\lower.5ex\hbox{$<$}\ }
\def\roughly#1{\raise.3ex\hbox{$#1$\kern-.75em\lower1ex\hbox{$\sim$}}}

 at 10truept
\def \dettwo {\phi^{[f}_f \phi^{g]}_g}
\def \detthree {\phi^{[f}_f \phi^g_g \phi ^{h]}_h}
\def \eomphi {\varepsilon^{\phi}}
\def \eomg {{\cal E}^{ab}}
\def \Eomg {\varepsilon^{ab}}

\def\d2{{(\hat D \pi)^2 }}

\def \n {{\nabla}}

\newcommand{\beq}{\begin{equation}}
\newcommand{\eeq}{\end{equation}}
\newcommand{\bea}{\begin{eqnarray}}
\newcommand{\eea}{\end{eqnarray}}
\newcommand{\beqa}{\begin{eqnarray}}
\newcommand{\eeqa}{\end{eqnarray}}

\begin{document}

\title{Boundary Terms and Junction Conditions for Generalized Scalar-Tensor Theories}

%\author{Antonio Padilla} 
%\email[]{antonio.padilla@nottingham.ac.uk}
\author{Antonio Padilla}
\email[]{antonio.padilla@nottingham.ac.uk}
\author{Vishagan Sivanesan}
\email[]{ppxvs@nottingham.ac.uk}
\affiliation{School of Physics and Astronomy ,University of Nottingham, Nottingham NG7 2RD, UK} 
\date{\today}

\begin{abstract}
We compute the boundary terms and junction conditions for Horndeski's panoptic class of scalar-tensor theories, and write the bulk and boundary equations of motion in explicitly second order form. We consider a number of special subclasses, including galileon theories, and present the corresponding formulae. Our analysis opens up  of the possibility of studying tunnelling between vacua in generalized scalar-tensor theories, and braneworld dynamics. The latter follows because our results are independent of spacetime dimension.
\end{abstract}

%\pacs{}
%\keywords{}

\maketitle

\section{Introduction}
The suggestion that the gravitational force  might contain an additional scalar component dates back to Kaluza and Klein's attempts to unify gravity with electromagnetism using five dimensional General Relativity \cite{Kaluza,Klein}, with the scalar field corresponding to  fluctuations in the size of the fifth dimension.  Scalar-tensor theories of gravitation were considered in their own right, first by Scherrer as early as 1941 \cite{Scherrer}, then independently by Jordan \cite{Jordan} and Thiry \cite{Thiry}\footnote{We thank Stanley Deser for enlightening us on the history of scalar-tensor theories. See \cite{history} for an historical overview.},  and most notably by Brans and Dicke \cite{BD}.  Such theories have received plenty of interest over the years,  especially within the cosmology community. This ranges from  early thoughts on  Dirac's large number hypothesis \cite{Dirac,Jordan} to recent  attempts to account for dark energy using modified gravity (for a review, see \cite{review}).   Current  interest is motivated, in part, by string theory, and the plethora of scalar fields that arise from string compactifications \cite{Grana}.  

Given the possible applications to cosmology, it is no surprise that the {\it most general} scalar-tensor theory was formulated by Horndeski in 1974 \cite{horny}. What is surprising is that this theory was forgotten about until very recently, where it was resurrected in \cite{fab4}, and discovered independently in \cite{dgsz}.  Horndeski's theory is the ``most general" scalar-tensor theory up to the requirement of second order field equations in four dimensions. Higher order field equations can be interpreted as propagating extra fields, and in any event, they typically suffer from the Ostrogradski instability \cite{ostro}. Here we will work with the DGSZ formulation \cite{dgsz} of Horndeski's theory, as it is more aesthetic and is valid in any number of dimensions\footnote{In four dimensions, the Horndeski and DGSZ actions were shown to be equivalent \cite{Kob}, and given Horndeski's proof, we know this to be the most general scalar-tensor theory  admitting second order field equations. In higher dimensions the DGSZ action is known to yield second order field equations, but it is not the most general theory.}. This is given by
 \begin{multline}  \label{acthorny}
S[g_{ab}, \phi]=\int_{\cal M}  k(\phi, X)-G_3(\phi, X) \Box \phi+G_4 (\phi, X)R+G_{4 X} \nabla^a_{[a} \phi  \nabla^b_{b]} \phi
+G_5(\phi, X) G_{ab}\nabla^a \nabla^b \phi-\frac{G_{5 X}}{6}  \nabla^a_{[a} \phi  \nabla^b_{b} \phi \nabla^c_{c]} \phi
 \end{multline}
 where $X=-\frac{1}{2} (\nabla \phi)^2$, and {\bf the antisymmetrisation does not include the usual factor of $1/n!$}. The covariant measure on the manifold is omitted for brevity.
 
As they stand, neither the original Horndeski action \cite{horny}, nor the recent reformulation  \cite{dgsz} given above admit a well defined variational principle on a manifold with a boundary.  This is problematic if one wishes to apply (Euclidean) path-integral methods to Horndeski's theory, or if one wishes to consider the dynamics of domain walls configurations. The same is true, of course, of the Einstein-Hilbert action, where the Gibbons-Hawking boundary term \cite{York,GH} is added such that the full theory can be extremised with Dirichilet boundary conditions on the spacetime metric. In this paper, we derive the analogue of the Gibbons-Hawking boundary term for Horndeski's theory. 

Armed with a well defined action, we can derive the junction conditions across a co-dimension one brane, or domain wall, embedded within the manifold. This leads to the analogue of the Israel junction conditions \cite{Israel} in Horndeski's theory, and opens up the possiblity of studying plenty of new physics from  bubble nucleation to  braneworld dynamics.  Our derivation makes  use of the standard technique of treating the brane as the common boundary of the bulk geometry on either side of the brane. The methods used for deriving the boundary terms and junction conditions will be described in more detail in section \ref{sec:method}, where we will explicitly present the relevant calculation for the first two terms in (\ref{acthorny}). The boundary terms and junction conditions for the full theory will be presented in section \ref{sec:results}.  In section \ref{sec:examples} we will discuss some special cases such as  Brans-Dicke gravity \cite{BD}, flat space galileon \cite{gal} and covariant galileon \cite{covgal} theory. We will conclude in section \ref{sec:conc}.

\section{Methodology} \label{sec:method}
Let us briefly outline the methodology we used in deriving the results that will be presented in the next section.  Consider  the incomplete action (\ref{acthorny}) defined on a manifold $\cal M$ with boundary $\del \cal M$. The boundary may be spacelike ($s=-1$) or timelike ($s=+1$). We  begin by computing the variation of  (\ref{acthorny}) keeping track of all surface terms. The result is 
\be
\delta S[g_{ab}, \phi]=\int_{\cal M} \Eomg \delta g_{ab} +\eomphi \delta \phi+\int_{\del \cal M} X^{ij} \delta h_{ij} +X^\phi \delta \phi+Y^{ij} \delta( h_{ij, n}) +Y^\phi \delta \phi_n
\ee
where $\Eomg$ and $\eomphi$ are the equations of motion.  We are using bulk coordinates $x^a$, and boundary coordinates $\xi^i$, and  we may think of the boundary as an embedding $x^a=X^a(\xi)$. This defines tangent vectors $\del_i X^a$, each of which is orthogonal to the unit outward point normal $n^a$. The induced metric on the boundary is defined as
\be
h_{ij}= \del_i X^a\del_j X^b g_{ab}|_{\del \cal M}
\ee
This can also be identified with the projector on to the boundary, which we denote $h_{ab}=g_{ab}-s n_a n_b$, where  $s=g_{ab} n^a n^b$.

Dirichilet boundary conditions require that $\delta \phi$ and $\delta h_{ij}$ vanish on $\del \cal M$, so the boundary  terms $X^{ij} \delta h_{ij}$ and $X^\phi \delta \phi$  are not considered problematic. The same cannot be said of the remaining boundary terms $Y^{ij} \delta( h_{ij, n})$ and $Y^\phi \delta \phi_n$. $\phi_n=n^a \del_a \phi|_{\del \cal M}$ is the normal derivative to the scalar on the boundary, and its variation is not necessarily vanishing. Similarly, $ h_{ij, n} =\del_i X^a\del_j X^b n^c \del_c g_{ab}|_{\del \cal M}$, which is the normal derivative to the metric on the boundary. These troublesome boundary terms are present because the DGSZ action (\ref{acthorny}) contains terms with second order derivatives.

To fix this problem, we must add a boundary term $B[h_{ij}, \phi, h_{ij,n}, \phi_n]$ whose variation cancels off the troublesome contributions described above. In other words, we must choose $B$ such that
\be
\delta B[h_{ij}, \phi, h_{ij,n}, \phi_n]=\int_{\del \cal M} Z^{ij} \delta h_{ij} +Z^\phi \delta \phi-Y^{ij} \delta( h_{ij, n}) -Y^\phi \delta \phi_n
\ee
It then follows that the total action $S_{total}=S+B$ admits a well defined variational principle, since 
\be
\delta S_{total}=\int_{\cal M} \Eomg\delta g_{ab} +\eomphi \delta \phi+\int_{\del \cal M} J^{ij} \delta h_{ij} +J^\phi\delta \phi
\ee
where $J^{ij}=X^{ij}+Z^{ij}$ and $J^\phi=X^\phi+Z^\phi$. Now, it is immediately clear that the choice of $B$ is not unique: if $B$ is a good boundary term, then so is $B+\eta[h_{ij}, \phi]$, since the variation of $\eta$ acts only to renormalise $Z^{ij}$ and $Z^\phi$. The same is of course true for the Gibbons-Hawking term in General Relativity. To eliminate this ambiguity, we impose a minimal construction, requiring that $B \to 0$ as both $h_{ij,n} \to 0$ and $ \phi_n \to 0$.

The junction conditions across a domain wall, $\Sigma \in {\cal M}$, can now  be derived  in one of two ways.  The first is to treat the wall as a delta-function source in the field equations. A completely equivalent approach, and the one we will adopt here, is  to note that the wall splits the manifold ${\cal M}$ into two manifolds, ${\cal M}_+$ and ${\cal M}_-$, and is   treated as the common boundary to each.  Of course, this statement neglects the contribution of boundary components far away from the wall, since they play no role here. The action describing the system is given by
\be
S_{DW}=S_{total}^++S^-_{total}+S_{\Sigma}
\ee
where $S_{total}^\pm$ is the total action defined on ${\cal M}_\pm$ with boundary $\del {\cal M}_\pm$. Variation of the full action yields
\begin{multline}
\delta S_{DW}=\int_{{\cal M}_+} \Eomg \delta g_{ab} +\eomphi \delta \phi+\int_{\del{ \cal M}_+} J^{ij} \delta h_{ij} +J^\phi\delta \phi \\+\int_{{\cal M}_-} \Eomg \delta g_{ab} +\eomphi \delta \phi+\int_{\del{ \cal M}_-} J^{ij} \delta h_{ij} +J^\phi\delta \phi+\int_{\Sigma} \frac{1}{\sqrt{-h}}\frac{\delta S_{\Sigma}}{\delta h_{ij}} \delta h_{ij} +\frac{1}{\sqrt{-h}} \frac{\delta S_{\Sigma}}{\delta \phi} \delta \phi
\end{multline}
Now because of the orientation, it is clear that $\int_{\del{ \cal M}_+}=-\int_{\del{ \cal M}_-}=\int_\Sigma$. It follows that
\be
\delta S_{DW}=\int_{{\cal M}_+ \cup {\cal M}_- }\Eomg \delta g_{ab} +\eomphi  \delta \phi+\int_{\Sigma}\left(\Delta J^{ij} + \frac{1}{\sqrt{-h}}\frac{\delta S_{\Sigma}}{\delta h_{ij}}\right)\delta h_{ij} +\left(\Delta J^\phi+\frac{1}{\sqrt{-h}}\frac{\delta S_{\Sigma}}{\delta \phi} \right)\delta \phi
\ee
where $\Delta Q=Q_{\del {\cal M}_+}-Q_{\del {\cal M}_-}$. The resulting junction conditions are given by the continuity relations $\Delta h_{ij}=\Delta \phi=0$ and the analogue of the Israel equations,
\be \label{israel}
\Delta J^{ij} =-\frac{1}{\sqrt{-h}} \frac{\delta S_{\Sigma}}{\delta h_{ij}}, \qquad \Delta J^\phi=-\frac{1}{\sqrt{-h}}\frac{\delta S_{\Sigma}}{\delta \phi}
\ee
Note that the continuity relations ensure that equations (\ref{israel}) are invariant under $B \to B+\eta[h_{ij}, \phi]$.

We shall now demonstrate explicitly how this methodology was applied  to the first two terms in (\ref{acthorny}). We begin with the $k$-essence term \cite{kessence}, $S_k=\int_{\cal M} k(\phi, X)$.  Variation yields
\be
\delta S_k=\int_{\cal M} \frac{1}{2} \left[k_X \nabla^a \phi \nabla^b \phi+k g^{ab}\right] \delta g_{ab}+\left[k_\phi+ \nabla_a (k_X \nabla^a \phi)\right] \delta \phi +\int_{\del \cal M} -k_X \phi_n \delta \phi
\ee
Because there were no second derivatives in $S_k$ this piece of the action is already well defined, and there is no need to add a boundary term. The contribution to the equations of motion and junction conditions can be immediately read off:
\ba
 \Eomg_k=\frac{1}{2} \left[k_X \nabla^a \phi \nabla^b \phi+k g^{ab}\right]  &\qquad& \eomphi_k=k_\phi+ \nabla_a (k_X \nabla^a \phi) \\
  J_k^{ij}=0 &\qquad&  J_k^\phi =\Delta \left[-k_X \phi_n\right]
\ea
Next we consider the second term in the DGSZ action (\ref{acthorny}), $S_3=-\int_{\cal M} G_3(\phi, X) \Box \phi$. We shall perform the variation with respect to $\phi$ and $g_{ab}$ separately. Starting with the $\phi$ variation, we find, 
\ba
\delta_{\phi}S_3 &=& \int_{\cal M} \bigg \{ -G_{3\phi} \Box \phi - (G_{3X} \phi^a)_{;a} \Box \phi - G_{3X} \phi_b \n^b \n^c \n_c \phi - \Box G_3 \bigg \}\delta \phi \\\nonumber
&& + \int_{\del \cal M} \left[ G_{3X} \phi_n \Box \phi   + G_{3n}\right] \delta \phi - G_3 \delta \phi_n 
\ea
where $G_{3n}\equiv n^a \nabla_a G_3$. The boundary terms contain the problematic contribution from $\delta \phi_n$. To cancel this off, we add the following:
\be
B_3=\int_{\del \cal M} F_3(\phi, Y,  \phi_n) 
\ee
where $Y=-\frac{1}{2}h^{ij} \del_i \phi \del_j\phi$  is the boundary analogue of $X$, and 
\be
F_3(\phi, Y, \phi_n)= \int^{\phi_n}_0 dx\,\, G_3\left(\phi, Y-\frac{1}{2}s x^2  \right) 
\ee
To see that this works, we note that 
\be
\delta_\phi B_{3} = \int_B G_3 \delta \phi_n  + \left[F_{3\phi}  - (F_{3Y} \phi^i)_{;i}\right] \delta \phi
\ee
The $\phi$ variation of the completed action is well behaved, and yields
\be
\delta_\phi (S_3+B_3)=\int_{\cal M} \eomphi_3 \delta \phi +\int_{\del \cal M} J^\phi_3 \delta \phi
\ee
where
\ba
\eomphi_3 &=&  -G_{3\phi}\Box \phi -\left(G_{3X}\phi^b\right )_{;b} \Box \phi + G_{3X}R_{ab} \phi^a \phi^b +(G_{3X}\phi^a)^{;b} \phi_{ab} -\left ( G_{3\phi} \phi_a \right)^{;a} \\
J_3^\phi &=& G_{3X}C \phi_n + G_{3\phi} \phi_n +G_{3X} K_{ij} \phi^{i} \phi ^{j} - F_{3YY}\phi^{i}\phi^{j}\phi_{ij} + F_{3Y} \bar \Box \phi  + F_{3\phi} +F_{3Y\phi}\phi_i\phi^i
\ea
A few comments are in order here. In arriving at the expression for $\eomphi_3$ we have eliminated the apparent third derivative terms using the Riemann identify, giving
\be
-G_{3X} \phi^b \n_b\n^c \n_c\phi  - \Box G_3 = G_{3X} R_{ab} \phi^a \phi^b + (G_{3X} \phi^b)^{;a} \phi_{ab} - (G_{3\phi} \phi_a)^{;a}
\ee
This serves as a good check of our calculation as we know that the equations of motion are second order.   Note that we sometimes denote covariant derivatives using superscripts and subscripts, ie $\phi_a=\nabla_a \phi, ~\phi^a=\nabla^a \phi$ etc. Covariant derivatives along the normal direction attain the super/subscript $n$, ie $\phi_n=n^a \phi_a, \phi_{nn}=n^a n^b \phi_{ab}$.

Similarly, the final expression for $J_3^\phi$ has made use of the following identity
\be
G_{3n} \equiv n^a \n_a G_3 = G_{3\phi} \phi_n -s G_{3X} \phi_{nn} - G_{3X} \phi_{ni} \phi^i + G_{3X}K_{ij} \phi^{ij}
\ee
where $K_{ij}$ is the extrinsic curvature of the boundary, defined as the Lie derivative of the induced metric with respect to the normal
\be
K_{ij}=\frac{1}{2} {\cal L}_n h_{ij}
\ee
We also introduce the covariant derivative on the boundary, $\bar D_i$, which we will sometimes denote using superscripts and subscripts, as with the bulk covariant derivative, ie  $\phi_i=\bar D_i \phi, ~\phi^i=\bar D^i \phi$. The covariant d'Alembertian on the boundary is written as $\bar \Box=\bar D_i \bar D^i$, while the boundary scalar $C$ is defined as the trace $C=h^{ij} C_{ij}$, where
\be
C_{ij}=\bar D_i \bar D_j \phi+s\phi_n K_{ij} 
\ee
In other words $C=\bar \Box \phi +s \phi_n K$ where $K=h^{ij} K_{ij}$. Further details of the useful formulae used in our derivations can be found in appendix \ref{sec:useful}. Once again we note that $J^\phi_3$ contains no more than second derivatives along the boundary, and first derivatives along the normal. This is to be expected for a second order system in the bulk.

We now consider the variation of $S_3$ with respect to the metric $g_{ab}$. This gives,
\be
\delta_gS_3 = \int_{\cal M}-\frac{1}{2}\left[G_3 \Box \phi g^{ab} + G_{3X} \Box \phi \phi^a \phi^b  + G_3^{;a}\phi ^{b}+ G_3^{;b}\phi^{a} - \left ( G_3 \phi^c \right) _{;c} g^{ab} \right] \delta g_{ab} \\\nonumber +\int_{\del \cal M}  -\frac{1}{2}G_3 \phi_n h^{ij} \delta{h_{ij}}    
\ee
Although the metric variation does not lead to any troublesome boundary terms, we must account for any additional contributions coming from $B_3$. The metric variation of $B_3$ yields
\be
\delta_gB_{3} = \int_{\del \cal M} \frac{1}{2}\left[ F_3 h^{ij} + F_{3Y} \phi^i \phi^j \right ] \delta h_{ij}
\ee
It follows that 
\be
\delta_g (S_3+B_3)=\int_{\cal M} \Eomg_3 \delta g_{ab} +\int_{\del \cal M} J^{ij}_3 \delta h_{ij}
\ee
where
\ba
\Eomg_3 &=& -\frac{1}{2}\left[G_3 \Box \phi g^{ab} + G_{3X} \Box \phi \phi^a \phi^b + G_3^{;a}\phi ^{b}+ G_3^{;b}\phi^{a} - \left ( G_3 \phi^c \right) _{;c} g^{ab} \right] \\
J^{ij}_3 &=&  \frac{1}{2} \left[ F_3 h^{ij} + F_{3Y} \phi^i \phi^j - G_3\phi_n h^{ij} \right ]
\ea
Again we see that the metric equations of motion are second order in the bulk, and the junctions conditions  contain no more than second derivatives along the boundary, and first derivatives along the normal.

Analogous calculations were applied to the remaining terms in the DGSZ action, which we denote
\be
S_4=\int_{\cal M} G_4 (\phi, X)R+G_{4 X} \nabla^a_{[a} \phi  \nabla^b_{b]} \phi, \qquad S_5= \int_{\cal M}
G_5(\phi, X) G_{ab}\nabla^a \nabla^b \phi-\frac{G_{5 X}}{6}  \nabla^a_{[a} \phi  \nabla^b_{b} \phi \nabla^c_{c]} \phi
\ee
However, the algebra is extremely long  so we shall not present it here, being content to present the results in the next section. Further details may be found in the forthcoming PhD thesis \cite{vishthesis}.
\section{Boundary terms and junction conditions for Horndeski theory} \label{sec:results}
In this section we shall simply quote the results of lengthy calculations, as described in the previous section. Our starting point is the DGSZ action for a general scalar-tensor theory \cite{dgsz}, which is equivalent to Horndeski's original theory \cite{horny} in four dimensions \cite{Kob}. Let us repeat the form of this action in order to make this section self-contained:
\begin{multline}  \label{acthorny1}
S[g_{ab}, \phi]=\int_{\cal M}  k(\phi, X)-G_3(\phi, X) \Box \phi+G_4 (\phi, X)R+G_{4 X} \nabla^a_{[a} \phi  \nabla^b_{b]} \phi
+G_5(\phi, X) G_{ab}\nabla^a \nabla^b \phi-\frac{G_{5 X}}{6}  \nabla^a_{[a} \phi  \nabla^b_{b} \phi \nabla^c_{c]} \phi
 \end{multline}
 where $X=-\frac{1}{2} (\nabla \phi)^2$. Recall that the antisymmetrisation does not include the usual factor of $1/n!$, and that the covariant measure on the manifold is omitted for brevity. In order to admit a well defined variational principle under Dirichilet boundary conditions, this action must be supplemented by the following boundary term
 \be
 B[h_{ij}, \phi, h_{ij,n}, \phi_n]=\sum_{\alpha=3}^5 B_\alpha[h_{ij}, \phi, h_{ij,n}, \phi_n]
\ee
where
\ba
B_{3} &=& \int_{\del \cal M} F_3 \\\nonumber
B_{4} &=& \int_{\del \cal M} 2(G_4 K - F_{4Y} \phi_i^i) \\\nonumber
B_{5} &=& \int_{\del \cal M} -\frac{1}{2} sG_5 K^{[i}_i K^{j]}_j \phi_n - G_5 \phi^{[i}_iK^{j]}_j + \frac{1}{2}{\bar R}F_5 + \frac{1}{2}F_{5Y} \phi^{[i}_i\phi^{j]}_j 
\ea
Here we define 
\be
F_\alpha (\phi, Y, \phi_n)= \int^{\phi_n}_0 dx\,\, G_\alpha\left(\phi, Y-\frac{1}{2}s x^2 \right), \qquad Y = -\frac{1}{2} \phi_i \phi^i
\ee
from which it follows that $\frac{\del F_\alpha}{\del \phi_n}=G_\alpha$.  Note that any curvature terms with an ``overbar" correspond to {\it boundary} curvatures, eg $\bar R_{ijkl}$ is the boundary Riemann tensor, $\bar G_{ij}$ is the boundary Einstein tensor, $\bar R$ is the boundary  Ricci scalar,  etc etc. Of course, if the ``overbar" is absent, it corresponds to a bulk curvature.

Variation of the full action, $S_{total}=S+B$ now yields, 
\be
\delta S_{total}=\int_{\cal M} \Eomg\delta g_{ab} +\eomphi \delta \phi+\int_{\del \cal M} J^{ij} \delta h_{ij} +J^\phi\delta \phi
\ee
where the bulk equations of motion are given by
\be
\Eomg = \frac{1}{2}( \eomg+{\cal E}^{ba}), \qquad   \eomg= \eomg_k+\sum_{\alpha=3}^5  \eomg_\alpha;  \qquad\qquad  \eomphi=\eomphi_k+\sum_{\alpha=3}^5 \eomphi_\alpha
\ee
with
\ba
\eomg_k &=& \frac{1}{2}(k_X \phi^a \phi^b +kg^{ab})\\
\eomg_3 &=& -\frac{1}{2}\left[G_3 \Box \phi g^{ab} + G_{3X} \Box \phi \phi^a \phi^b + 2 G_3^{;a}\phi ^{b} - \left ( G_3 \phi^c \right) _{;c} g^{ab} \right]\\
\eomg_4 &=& \frac{1}{2}\big(g^{ab}G_{4X}\dettwo + G_{4X}R\phi^a \phi^b - 2 G_4 G^{ab} + G_{4XX}\dettwo \phi^a \phi^b \big) - {(G_{4\phi} \phi_c)^{;[c}g^{a]b} }\nonumber \\
&&+ (G_{4X} \phi_d)^{;[c} g^{a]b} \phi^d_c  + G_{4X}\phi^d g^{a[b}R^{c]}_{\,dce} \phi^e + 2 G_{4X}^{;[a}\phi^{c]}_c \phi^b -2 G_{4X} R^{a}_c \phi^{b} \phi^c  - {\color{green} } (G_{4X}\phi^d)_{;d} g^{a[b}\phi^{c]}_c \\
\eomg_5 &=& \frac{1}{2}\bigg[ 
G_5(R^{ab} \Box \phi- R \phi^{ab}) - 4G_5 G^a_c \phi^{cb} + 2(G_5 \phi^a)_{;d}G^{bd} - (G_5 \phi^c)_{;c} G^{ab} - G_{5X}^{;[a}\phi^c_c \phi^{d]}_d \phi^b \nonumber  \\\nonumber
&&  +\frac{1}{2} G_{5X;d} \phi^d g^{a[b} \phi^c_c \phi^{e]}_e + \frac{1}{2} G_{5X} (\Box \phi) g^{a[b} \phi^c_c \phi^{d]}_d +G_{5X} G_{cd} \phi^{cd} \phi^a \phi^b  - \frac{1}{6} G_{5XX} \detthree \phi^a \phi^b\\\nonumber
&& g^{a[b} \phi^c_c \nabla^{d]}(G_{5\phi} \phi_d) - g^{a[b} \phi^c_c \nabla^{d]}(G_{5X}\phi_e) \phi^e_d  - G_{5X} g^{a[b} \phi^c_c R^{d]}{}_{edf} \phi^e \phi^f -G_{5X} \phi^a \phi^{[c}_c R^{bd]}{}{}_{de} \phi^e \\
&& + 2G_5^{;a} R^b_c\phi^c -2G_{5;c}R^{cabd}\phi_d  - 2 G_{5;c} R^{cd} \phi_d  g^{ab} + 2G_5 R^{a}_d \phi^{bd}- \frac{1}{2} G_5 R \Box \phi g^{ab} -\frac{G_{5X}}{6} g^{ab} \detthree \bigg]
\ea
and
\ba
\eomphi_k &=& k_\phi + (k_X \phi_a)^{;a}\\
\eomphi_3 &=& -G_{3\phi}\Box \phi -\left(G_{3X}\phi^b\right )_{;b} \Box \phi + G_{3X}R_{ab} \phi^a \phi^b +(G_{3X}\phi^a)^{;b} \phi_{ab} -\left ( G_{3\phi} \phi_a \right)^{;a}\\
\eomphi_4 &=&  G_{4\phi}R + (G_{4X}\phi_a)^{;a} R + G_{4X\phi}\dettwo +(G_{4XX}\phi_a)^{;a}\dettwo - 2G_{4XX} \phi^{[b}_bR^{a]}_{\,cad}\phi^c \phi^d + 2(G_{4X\phi}\phi_a)^{[;a}\phi^{b]}_b \nonumber \\ 
&&- 2(G_{4XX}\phi_c)^{[;a}\phi^{b]}_b \phi^c_a - 4R_{ab}G_{4X}^{;a}\phi^b {-2G_{4X} R_{ab}\phi^{ab}} \\
\eomphi_5 &=& G_{5\phi} G_{ab} \phi^{ab} + (G_{5X}\phi_c)^{;c} G_{ab} \phi^{ab} - \frac{1}{6} G_{5X\phi} \detthree - \frac{1}{6}(G_{5XX}\phi_c)^{;c} \detthree  \nonumber \\\nonumber
&&+(G_{5\phi}\phi^a)^{;b}G_{ab} - (G_{5X}\phi_c)^{;b} \phi^{ac} G_{ab} + G_{5X}R_{abcd}G^{ad}\phi^b \phi^c -\frac{1}{2} (G_{5X\phi}\phi_a)^{;[a}\phi^b_b\phi^{c]}_c \\\nonumber
&&+\frac{1}{2} (G_{5XX}\phi^d)^{;[a}\phi^b_b\phi^{c]}_c \phi_{ad} + \frac{1}{2} G_{5XX}\phi^d \phi^e \phi^{[a}_a\phi^b_bR^{c]}{}_{dce} -G_{5X;a} \phi^{[b}_b R^{ac]}{}{}_{cd}\phi^d \\
&& -\frac{1}{2}G_{5X} R^{[ab}{}{}_{bd} R_a{}^{c]}{}{}_{ce} \phi^d \phi^e  - G_{5X}\phi^{[a}_d \phi^b_b R_a{}^{c]}{}{}_c{}^d 
\ea
Note that we have written the bulk equations of motion in a form that is explicitly second order, something that has yet to appear in the literature, as far as we are aware.

As explained in the previous section, the junction conditions (\ref{israel}) can be obtained  from the boundary equations of motion, which are given by
\be
J^{ij}=\frac{1}{2} ( {\cal J}^{ij}+{\cal J}^{ji}), \qquad {\cal J}^{ij}=\sum_{\alpha=3}^5  {\cal J}^{ij}_\alpha; \qquad \qquad  J^\phi_\alpha=J^\phi_k+\sum_{\alpha=3}^5  J^{\phi}_\alpha
\ee
with
\ba
{\cal J}_3^{ij} &=& \frac{1}{2} \left [ F_3 h^{ij} + F_{3Y} \phi^{i}\phi^{j} -G_3\phi_n h^{ij} \right] \\
{\cal J}_4^{ij}&=& -G_4(K^{ij}-Kh^{ij}) + G_{4\phi} \phi_n h^{ij} - G_{4X} \phi^k \phi_{nk} h^{ij} + G_{4X}\phi^k\phi^l K_{kl} h^{ij} + 2s G_{4X} B^i \phi^j  \nonumber \\
&&+ G_{4X} \phi_n h^{i[j}C^{k]}_k +G_{4X} K \phi^i \phi^j - F_{4YY} \phi^i \phi^j \bar \Box \phi- 2 F_{4Y}^{;i} \phi^j  + F_{4Y;k} \phi^k h^{ij}  \\
{\cal J}_5^{ij}&=&\frac{1}{2} \bigg[{-\frac{1}{2} s G_{5X}K^{[k}_kK^{l]}_l \phi^i \phi^j  \phi_n}- G_{5X} \phi^{[k}_kK^{l]}_l \phi^i \phi^j -2G_5^{;[i}K^{k]}_k \phi^j \nonumber \\
\nonumber
&&+ G_{5;k} \phi^k h^{i[j} K^{l]}_l - F_5 {\bar G}^{ij} + \frac{1}{2}F_{5Y}{\bar R} \phi^i \phi^j + \frac{1}{2} F_{5YY} \phi^{[k}_k \phi^{l]}_l \phi^i \phi^j + 2 F_{5Y}^{[;i}\phi^{k]}_k \phi^j \\\nonumber
&&+2sG_{5;k}B^kh^{ij} -2s G_5^{;i} B^j - \phi_{nk}G_5^{;[k}h^{i]j} + (F_{5Y}\phi_k)^{[;l}h^{i]j} \phi^k_l - (F_{5Y}\phi^k)_{;k} h^{i[j}\phi^{l]}_l + F_{5Y}\phi^k {\bar R}_{klm} {}{}{}^{[l}h^{i]j}\phi^m \\\nonumber
&&+ G_5 \phi_n {\bar G}^{ij} - G_{5\phi}\phi_n h^{i[j} C^{k]}_k + G_{5X} \phi_{nk}\phi^k h^{i[j}C^{l]}_l -G_{5X} K_{kl} \phi^k \phi^l h^{i[j} C^{m]}_m \\
&&-2s G_{5X} \phi^i B^{[j} C^{k]}_k +s G_{5X} \phi_n h^{i[j} B^{k]}B_k - \frac{1}{2}G_{5X}\phi_n h^{i[j} C^k_k C^{l]}_l + \frac{1}{2} F_{5Y} \phi^{[k}_k \phi^{l]}_l h^{ij}{-2F_{5Y} \bar R^i_k \phi^j \phi^k} \bigg]
\ea
and
\ba
J_k^{\phi} &=& -k_X \phi_n \\
J_3^{\phi}&=& G_{3X}C \phi_n + G_{3\phi} \phi_n +G_{3X} K_{ij} \phi^{i} \phi ^{j} - F_{3YY}\phi^{i}\phi^{j}\phi_{ij} + F_{3Y} \bar \Box \phi  + F_{3\phi} +F_{3Y\phi}\phi_i\phi^i\\
J_4^{\phi}&=&  -G_{4X} \phi_n ({\bar R} -s K^{[i}_iK^{j]}_j) - G_{4XX}\phi_n[-2s  B^iB_i + C^{[i}_iC^{j]}_j] +4s G_{4X;i}B^i -2 G_{4X\phi} \phi_n C + 2 G_{4XX}C \phi_{ni}\phi^i \nonumber \\
&& -2G_{4XX}CK_{ij}\phi^i\phi^j - 2 G_{4X}K_{ij}\phi^{ij}+2G_{4\phi}K + 2(G_{4X}\phi_i)^{;i} K  - 2(F_{4YY}\phi_i)^{;i} \bar \Box \phi + 2F_{4YY}{\bar R}_{ij} \phi^i \phi^j - 2 G_{4X;i} \phi_n^i \nonumber  \\
&&+ 2 (F_{4YY}\phi^i)_{;j} \phi^j_i {- 2(F_{4Y\phi}\phi_i)^{;i} }-2 F_{4Y\phi} \bar \Box \phi \\
J_5^\phi &=& -s G_{5X;i}B^{[i}C^{j]}_j - G_{5X}C^{ij}\phi_n\left({\bar G}_{ij} -s \left[K K_{ij} - 2K_{ik}K^k_j -\frac{1}{2}h_{ij}(K^2 + K_{kl}K^{kl})\right]\right) - G_{5\phi} \phi^i (K_{ij}{}^{;j} - K_{;i}) \nonumber \\
&&+\frac{1}{2}(G_{5\phi}\phi_n -s G_{5X} \phi^i B_i)({\bar R} -s K^{[k}_kK^{l]}_l) + G_{5X} C^{ij}\phi_j(K_{ik}{}^{;k} - K_{;i}) + \frac{1}{6} G_{5XX} C^{[i}_iC^j_jC^{k]}_k \phi_n \nonumber  \\\nonumber
&&+ \frac{1}{2}G_{5\phi X} \phi_n C^{[i}_i C^{j]}_j - \frac{1}{2}s G_{5XX}\phi_iB^i C^{[k}_kC^{l]}_l -s G_{5\phi X} \phi_i B^{[i}C^{i]}_i +s G_{5XX} \phi_i C^i_j B^{[j}C^{k]}_k -G_{5X}C \phi^i (K_{ij}{}{}^{;j}-K_{;i})\\\nonumber
&& +s G_{5X}C \phi_n K_{ij}K^{ij} +s G_{5X} \phi^i B^j ({\bar R}_{ij} -s KK_{ij} +2s K_{ik}K^k_j) + G_{5X}C^{ij} \phi^{k}K_{i[k;j]} - G_{5X} B^i \phi^j K_{ik}K^k_j \\\nonumber
&&+s G_{5X}\phi_n K_{ik}K^{kj}C^i_j  - \frac{1}{2}s G_{5\phi} K^{[i}_iK^{j]}_j\phi_n - \frac{1}{2}s(G_{5X}K^{[i}_iK^{j]}_j\phi_n\phi_k)^{;k} - G_{5\phi} \phi^{[i}_i K^{j]}_j - (G_{5X}\phi_i)^{;i} \phi^{[j}_jK^{k]}_k\\\nonumber
&& - G_{5X}\phi^{[i}_i K^{j]}{}_{j;k}\phi^k + \frac{1}{2}(F_{5Y}\phi_i)^{;i} {\bar R} - (G_{5\phi}\phi_i)^{[;i}K^{j]}{}_j +(G_{5X}\phi^i)^{[;j}K^{k]}{}_k\phi_{ij} +s (G_{5X}\phi_n)^{[;i}K^{j]}_j\phi_{ni}\\\nonumber
&& -2 G_5^{[;i}K^{j]}{}_{j;i} + \frac{1}{2}(F_{5YY}\phi_i)^{;[i}\phi^j_j\phi^{k]}_k + G_{5X;i}\phi_n{}^{;[i}\phi^{j]}_j + G_{5X}(\phi^iK_{ij})^{[;j}C_k^{k]} - G_{5X} B^{[i}K^{j]}{}_{j;i}\phi_n \\\nonumber
&&-G_{5X} K^{[i}{}_i{\bar R}_{jk}{}{}^{k]l}\phi^j \phi_l + F_{5YY}\phi^{[i}_i{\bar R}_{jk}{}{}^{k]l} \phi^j \phi_l - F_{5Y}{\bar R}_{ij}\phi^{ij} - 2 F_{5Y;i} \phi_j {\bar R}^{ij} \\
&&+s G_{5X}{\bar R}_{ij}B^i \phi^j - G_{5;i}(K^{ij}{}{}_{;j} - K^{;i}) + (F_{5Y \phi} \phi_i)^{[;i}\phi^{j]}_j  +\frac{1}{2} F_{5\phi}{\bar R} + \frac{1}{2} F_{5Y\phi} \phi^{[i}_i \phi^{j]}_j - G_{5X}\phi_{ni} B^{[i}K^{j]}_j 
\ea
Here we recall that
\be \label{Cij}
C_{ij}=\bar D_i \bar D_j\phi +s\phi_n K_{ij}, \qquad C=h^{ij} C_{ij}=\bar \Box \phi+s\phi_n K
\ee
and we introduce the boundary vector
\be \label{Bi}
B_i=s\bar D_i \phi_n-s K_{ij} \bar D^j \phi
\ee
Note also that $\phi_{ni}=\bar D_i \phi_n$. The formulae for $J^{ij}$ and $J^\phi$ have been written so that they are explictly second order in boundary derivatives, and first order in normal derivatives.

\section{Examples} \label{sec:examples}
We shall now present the boundary terms and junction conditions for certain important subclasses of Horndeski's theory, specifically: General Relativity (as a check), Brans-Dicke gravity \cite{BD}, covariant galileon theory \cite{covgal}, and the original flat space galileon theory \cite{gal}. Of course, one can use the results of the previous section to infer the boundary terms and junction conditions for many other theories such as {\it the Fab Four} \cite{fab4}, DBI theories \cite{dbi}, conformal galileon \cite{gal}, KGB theories \cite{kgb} and  so on.

\subsection{General Relativity}
General Relativity is perhaps the most ``special" special case of Horndeski's theory, corresponding to the choice
$$G_4=\frac{1}{16\pi G}, \qquad k=G_3=G_5=0$$
so that the bulk action is given by the standard Einstein-Hilbert action
\be
S=\frac{1}{16\pi G} \int_{\cal M} R
\ee
and, as expected, the boundary term is given by the Gibbons-Hawking term \cite{York, GH}
\be
B=\frac{1}{8\pi G} \int_{\del \cal M} K
\ee
The bulk equations of motion are simply the Einstein tensor
\be
\Eomg=-\frac{1}{16\pi G} G^{ab}
\ee
while the boundary equations of motion take the form expected from the Israel junction conditions \cite{Israel}
\be
J^{ij}=-\frac{1}{16\pi G} \left(K^{ij}-K h^{ij}\right)
\ee

\subsection{Brans-Dicke theory}
Brans-Dicke theory \cite{BD} is the most well studied scalar-tensor theory, and corresponds to the choice
\be
k=\frac{\omega}{8\pi\phi}X \qquad G_4=\frac{\phi}{16\pi}, \qquad G_3=G_5=0
\ee
so that the bulk action is given by
\be
S=\frac{1}{16 \pi} \int_{\cal M} \phi R-w\frac{(\nabla \phi)^2}{\phi}
 \ee
 and the boundary term by
 \be
 B=\frac{1}{8\pi}\int_{\del \cal M} \phi K
 \ee
 The bulk equations of motion are the usual Brans-Dicke field equations
 \ba
 \Eomg &=& -\frac{1}{16 \pi}\left[ \phi G^{ab} + g^{ab} \Box \phi-\phi^{ab}- \frac{\omega}{ \phi}\left(Xg^{ab} + \phi^a \phi^b \right )\right] \\
 \eomphi &=&\frac{1}{16\pi}\left[R+2w\left(\frac{\Box \phi}{\phi} +\frac{X}{\phi^2}\right) \right]
 \ea
 while the boundary equations of motion are
 \ba
 J^{ij} &=& \frac{1}{16 \pi} \left[ -\phi \left(K^{ij} - Kh^{ij}\right) + \phi_n h^{ij} \right] \\
 J^\phi &=& \frac{1}{8\pi}\left[K -\frac{\omega}{\phi}\phi_n\right]
 \ea
 It is easy to check that these are consistent with the junction conditions presented in \cite{BDjunc}.
\subsection{Covariant galileon}
Covariant galileon theory \cite{covgal} was developed in order to couple the original galileon theory \cite{gal} to gravity without introducing any new higher derivatives. We will ignore the historical timeline and begin by discussing the covariant model because the flat space galileon is easily obtained by  decoupling the graviton. Static spherically symmetric thin shells for the covariant  galileon, up to cubic order, were studied in \cite{hirsute} in order to explore aspects of the Vainshtein mechanism \cite{vainsh}.  This suggests that the following formulae will ultimately lend themselves to undertstanding screening mechanisms in modified gravity.

The covariant galileon theory corresponds to the choice,
\be
k=c_2 X, \qquad
G_3= -c_3 X, \qquad 
G_4=\frac{1}{2}c_4X^2, \qquad
G_5= -3c_5X^2
\ee
where $c_i$ are constant.  This gives the bulk action, 
\ba
S= \int_{\cal M} c_2 X + c_3 X \Box \phi + c_4 X\left( \dettwo +\frac{1}{2} X R \right) +c_5 X \left(  \detthree -3X G_{ab}\phi^{ab} \right) 
\ea
and the boundary term
\begin{multline}
B=\int_{\del \cal M} \left(c_3+2c_4 \bar \Box \phi+3c_5 \dettwo \right) \phi_n \left(\frac{1}{6}s \phi_n^2 - Y \right)-\frac{3}{2} c_5 \bar R \phi_n \left[\left(\frac{1}{6}s \phi_n^2 - Y \right)^2+\frac{\phi_n^4}{45} \right] \\
+c_4 X^2 K +\frac{3}{2} c_5 X^2 \left( sK^{[i}_i K^{j]}_j \phi_n+2\phi^{[i}_i K^{j]}_j \right)
\end{multline}
where we recall that $Y=-\frac{1}{2} \phi_i \phi^i$ is the boundary analogue of $X$. The bulk equations of motion now give 
$
\Eomg=\frac{1}{2} (\eomg+{\cal E}^{ba})
$
with 
\ba\nonumber
\eomg &=& \frac{1}{2}c_2 ( \phi^a \phi^b +Xg^{ab})\\\nonumber
 &&+c_3\left[ \frac{1}{2} \Box \phi \left[\phi^a \phi^b+X g^{ab}\right] +X^{(;a}\phi ^{b)} -\frac{1}{2} \left ( X \phi^c \right) _{;c} g^{ab}\right] \\
&&+c_4\left[ \frac{1}{2}\big(g^{ab}X\dettwo + XR\phi^a \phi^b - X^2 G^{ab} + \dettwo \phi^a \phi^b \big) \right. \nonumber \\ \nonumber
&&\qquad \left. + (X \phi_d)^{;[c} g^{a]b} \phi^d_c  + X\phi^d g^{a[b}R^{c]}_{\,dce} \phi^e + { 2} X^{;[a}\phi^{c]}_c \phi^b -2 X R^{a}_c \phi^{b} \phi^c  - {\color{green} } (X\phi^d)_{;d} g^{a[b}\phi^{c]}_c \right] \\ \nonumber
&&+ \frac{1}{2}c_5\bigg[ 
-3X^2(R^{ab} \Box \phi- R \phi^{ab}) +12X^2 G^a_c \phi^{cb} - 6(X^2\phi^a)_{;d}G^{bd} +3(X^2 \phi^c)_{;c} G^{ab} +6X^{;[a}\phi^c_c \phi^{d]}_d \phi^b \nonumber  \\\nonumber
&&\qquad ~  -3X_{;d} \phi^d g^{a[b} \phi^c_c \phi^{e]}_e -3X \Box \phi g^{a[b} \phi^c_c \phi^{d]}_d -6X G_{cd} \phi^{cd} \phi^a \phi^b  + \detthree \phi^a \phi^b\\\nonumber
&& \qquad ~+6g^{a[b} \phi^c_c \nabla^{d]}(X\phi_e) \phi^e_d  +6X g^{a[b} \phi^c_c R^{d]}{}_{edf} \phi^e \phi^f +6X \phi^a \phi^{[c}_c R^{bd]}{}{}_{de} \phi^e  -6(X^2)^{;a} R^b_c\phi^c \\
&& \qquad  ~+6(X^2)_{;c}R^{cabd}\phi_d  +6(X^2)_{;c} R^{cd} \phi_d  g^{ab} -6X^2 R^{a}_d \phi^{bd} + \frac{3}{2} X^2 R \Box \phi g^{ab} +X g^{ab} \detthree \bigg]
\ea
and
\ba \nonumber
\eomphi &=& c_2 \Box \phi\\\nonumber
&&+c_3\left[  (\Box \phi)^2 -R_{ab} \phi^a \phi^b -\phi^{ab} \phi_{ab}\right] \\\nonumber
&&+c_4 \left[ (X\phi_a)^{;a} R  +\Box \phi \dettwo - 2 \phi^{[b}_bR^{a]}_{\,cad}\phi^c \phi^d - 2\phi_c^{[a}\phi^{b]}_b \phi^c_a - 4R_{ab}X^{;a}\phi^b  { -2c_4 XR_{ab}\phi^{ab}} \right] \\\nonumber
&&+c_5 \bigg [ -6(X\phi_c)^{;c} G_{ab} \phi^{ab} +\Box\phi  \detthree +6(X\phi_c)^{;b} \phi^{ac} G_{ab} -6XR_{abcd}G^{ad}\phi^b \phi^c -3\phi^{d;[a}\phi^b_b\phi^{c]}_c \phi_{ad}  \nonumber \\
&&\qquad~-3\phi^d \phi^e \phi^{[a}_a\phi^b_bR^{c]}{}_{dce} +6X_{;a} \phi^{[b}_b R^{ac]}{}{}_{cd}\phi^d +{ 3X R^{[ab}{}{}_{bd} R_a{}^{c]}{}{}_{ce} \phi^d \phi^e } +6X\phi^{[a}_d \phi^b_b R_a{}^{c]}{}{}_c{}^d \bigg]  \label{ephi-covgal}
\ea
We have checked the consistency of these equations with the corresponding formulae presented in \cite{covgal}\footnote{The formula for $\Eomg$ matches exactly, while the formulae for $\eomphi$ differ by a term proportional to  $XR^{abc}{}_d R_{abce} \phi^d \phi^e$. We believe that \cite{covgal} contains a typo and that the formula (\ref{ephi-covgal}) presented here is correct.}.  The boundary equations of motion for variation of the metric are given by  $J^{ij}=\frac{1}{2}\left({\cal J}^{ij}+{\cal J}^{ji}\right)$ with, 
\ba
{\cal J}^{ij} &=& \frac{1}{2}c_3 \left [ Z \phi_n h^{ij} -\phi_n  \phi^{i}\phi^{j} +X\phi_n h^{ij} \right]  \nonumber \\
&&+c_4\left[ -\frac{1}{2} X^2 (K^{ij}-Kh^{ij})  - X \phi^k \phi_{nk} h^{ij} + X\phi^k\phi^l K_{kl} h^{ij} + 2s X B^i \phi^j \right. \nonumber \\
&&\qquad  \left.+ X \phi_n h^{i[j}C^{k]}_k +X K \phi^i \phi^j - \phi_n  \phi^i \phi^j \bar \Box \phi+2 Z \phi_n ^{;i} \phi^j  -(Z \phi_n)_{;k}  \phi^k h^{ij}\right]  \nonumber \\
&&+\frac{1}{2}c_5 \bigg[{3s XK^{[k}_kK^{l]}_l \phi^i \phi^j  \phi_n}+ 6X\phi^{[k}_kK^{l]}_l \phi^i \phi^j +6(X^2)^{;[i}K^{k]}_k \phi^j \nonumber \\
\nonumber
&&\qquad -3 (X^2)_{;k} \phi^k h^{i[j} K^{l]}_l +3\phi_n \left(Z^2+\frac{\phi_n^4}{45} \right) {\bar G}^{ij} + 3 Z \phi_n {\bar R} \phi^i \phi^j -3 \phi_n \phi^{[k}_k \phi^{l]}_l \phi^i \phi^j + 12 (Z \phi_n)^{[;i}\phi^{k]}_k \phi^j  \nonumber \\\nonumber
&&\qquad -6s(X^2)_{;k}B^kh^{ij} +6s (X^2)^{;i} B^j +3 \phi_{nk}(X^2)^{;[k}h^{i]j} + 6( Z \phi_n \phi_k)^{[;l}h^{i]j} \phi^k_l - 6( Z \phi_n\phi^k)_{;k} h^{i[j}\phi^{l]}_l\\\nonumber
&& \qquad + 6 Z \phi_n \phi^k {\bar R}_{klm} {}{}{}^{[l}h^{i]j}\phi^m -3X^2 \phi_n {\bar G}^{ij} -6X \phi_{nk}\phi^k h^{i[j}C^{l]}_l +6X K_{kl} \phi^k \phi^l h^{i[j} C^{m]}_m \\
&&\qquad +12s X \phi^i B^{[j} C^{k]}_k -6s X \phi_n h^{i[j} B^{k]}B_k +3X\phi_n h^{i[j} C^k_k C^{l]}_l + 3 Z \phi_n  \phi^{[k}_k \phi^{l]}_l h^{ij}{-12 Z \phi_n \bar R^i_k \phi^j \phi^k} \bigg]
\ea
where $X=-\frac{1}{2}s \phi_n^2 +Y$,  $Z=\frac{1}{6}s \phi_n^2 - Y$. The boundary equations of motion for variation of the scalar, meanwhile, are given by 
\ba
J_{\phi} &=& - c_2 \phi_n 
+c_3 \left[ -C \phi_n -K_{ij} \phi^{i} \phi ^{j}-\phi_n \bar \Box \phi \right]  \nonumber\\
&&+c_4 
\Big [  -X \phi_n ({\bar R} -s K^{[i}_iK^{j]}_j) - \phi_n[-2s  B^iB_i + C^{[i}_iC^{j]}_j] +4s X_{;i}B^i  + 2 C \phi_{ni}\phi^i  \nonumber \\ 
&& \qquad -2CK_{ij}\phi^i\phi^j - 2 XK_{ij}\phi^{ij}+ 2(X\phi_i)^{;i} K  - 2(\phi_n \phi_i)^{;i}\bar \Box \phi + 2\phi_n {\bar R}_{ij} \phi^i \phi^j - 2 X_{;i} \phi_n^i + 2 (\phi_n \phi^i)_{;j} \phi^j_i  \Big ] \nonumber \\
&& +c_5 \Bigg[ 6s X_{;i}B^{[i}C^{j]}_j +6XC^{ij}\phi_n\left({\bar G}_{ij} -s \left[K K_{ij} - 2K_{ik}K^k_j -\frac{1}{2}h_{ij}(K^2 + K_{kl}K^{kl})\right]\right)  \nonumber \\
&&\qquad +3s X \phi^i B_i ({\bar R} -s K^{[k}_kK^{l]}_l) -6X C^{ij}\phi_j(K_{ik}{}^{;k} - K_{;i}) -C^{[i}_iC^j_jC^{k]}_k \phi_n \nonumber  \\\nonumber
&&\qquad +3s\phi_iB^i C^{[k}_kC^{l]}_l -6s  \phi_i C^i_j B^{[j}C^{k]}_k +6XC \phi^i (K_{ij}{}{}^{;j}-K_{;i})\\\nonumber
&&\qquad  -6s XC \phi_n K_{ij}K^{ij} -6sX \phi^i B^j ({\bar R}_{ij} -s KK_{ij} +s K_{ik}K^k_j) -6XC^{ij} \phi^{k}K_{i[k;j]}  \\\nonumber
&&\qquad -6sX\phi_n K_{ik}K^{kj}C^i_j   +3s(XK^{[i}_iK^{j]}_j\phi_n\phi_k)^{;k} +6 (X\phi_i)^{;i} \phi^{[j}_jK^{k]}_k\\\nonumber
&& \qquad +6X\phi^{[i}_i K^{j]}{}_{j;k}\phi^k + 3(Z \phi_n \phi_i)^{;i} {\bar R} -6(X\phi^i)^{[;j}K^{k]}{}_k\phi_{ij} -6s (X\phi_n)^{[;i}K^{j]}_j\phi_{ni}\\\nonumber
&&\qquad  +6 (X^2)^{[;i}K^{j]}{}_{j;i} -3(\phi_n \phi_i)^{;[i}\phi^j_j\phi^{k]}_k -6X_{;i}\phi_n{}^{;[i}\phi^{j]}_j -6X(\phi^iK_{ij})^{[;j}C_k^{k]}+6X B^{[i}K^{j]}{}_{j;i}\phi_n \\\nonumber
&&\qquad +6X K^{[i}{}_i{\bar R}_{jk}{}{}^{k]l}\phi^j \phi_l -6 \phi_n \phi^{[i}_i{\bar R}_{jk}{}{}^{k]l} \phi^j \phi_l - 6 Z \phi_n {\bar R}_{ij}\phi^{ij} - 12(Z\phi_n)_{;i} \phi_j {\bar R}^{ij} \\
&&\qquad -6s X{\bar R}_{ij}B^i \phi^j +3(X^2)_{;i}(K^{ij}{}{}_{;j} - K^{;i}) +6 X\phi_{ni} B^{[i}K^{j]}_j \Bigg]
\ea
\subsection{Galileon in flat space}
The original galileon theory \cite{gal} corresponds to a single  scalar field propagating in Minkowksi space, satisfying the ``galileon" symmetry $\phi \to \phi+b_\mu x^\mu+c$, where $b_\mu$ and $c$ are constants. We can obtain the equations of motion and boundary terms for this theory by taking the limit $g_{\mu\nu} \to \eta_{\mu\nu}$ of the covariant galileon theory. It follows that we recover the (by now) well known  galileon action in the bulk \cite{gal}, 
\ba
S= \int_{\cal M} c_2 X + c_3 X \Box \phi + c_4 X \dettwo +c_5 X \detthree 
\ea
The boundary terms do not simplify as much, and are given by
\begin{multline} \label{Bgal}
B=\int_{\del \cal M} \left(c_3+2c_4 \bar \Box \phi+3c_5 \dettwo \right) \phi_n \left(\frac{1}{6}s \phi_n^2 - Y \right)-\frac{3}{2} c_5 \bar R \phi_n \left[\left(\frac{1}{6}s \phi_n^2 - Y \right)^2+\frac{\phi_n^4}{45} \right] \\
+c_4 X^2 K +\frac{3}{2} c_5 X^2 \left( \bar R \phi_n+2\phi^{[i}_i K^{j]}_j \right)
\end{multline}
One might be puzzled by the presence of curvature terms in this expression.  However, even though the bulk geometry is flat, the same need not be true of the boundary if it corresponds to a non-trivial embedding. That is not to say that there is no simplication whatsoever. Because  the bulk is  flat, the Gauss-Codazzi relations lead to the following identities,
\ba
\bar R_{ijkl}&=&  s K_{k[i} K_{j]l} \nonumber \\\nonumber
0&=& \bar D^jK_{ij} - \bar D_i K\ea
We have   already used  the first of these in expressing  (\ref{Bgal}).
 
 The bulk equations of motion are the usual galileon equations \cite{gal}, 
  \be 
\eomphi = c_2 \Box \phi+c_3 \dettwo +c_4 \detthree +c_5 \phi^{[f}_f \phi^g_g \phi^h_{h} \phi^{l]}_l \label{ephi-gal}
\ee
while the boundary equations of motion are given by%
{
\ba
J_{\phi} &=& - c_2 \phi_n 
+c_3 \left[ -C \phi_n -K_{ij} \phi^{i} \phi ^{j}-\phi_n \bar \Box \phi \right]  \nonumber\\
&&+c_4 
\Big [ - \phi_n[-2s  B^iB_i + C^{[i}_iC^{j]}_j] +4s X_{;i}B^i  + 2 C \phi_{ni}\phi^i  \nonumber \\ 
&& \qquad -2CK_{ij}\phi^i\phi^j - 2 XK_{ij}\phi^{ij}+ 2(X\phi_i)^{;i} K  - 2(\phi_n \phi_i)^{;i} \bar \Box \phi + 2\phi_n {\bar R}_{ij} \phi^i \phi^j - 2 X_{;i} \phi_n^i + 2 (\phi_n \phi^i)_{;j} \phi^j_i  \Big ] \nonumber \\
&& +c_5 \Bigg[ 6s X_{;i}B^{[i}C^{j]}_j +6XC^{ij}\phi_n\left({\bar G}_{ij} -s \left[K K_{ij} - 2K_{ik}K^k_j -\frac{1}{2}h_{ij}(K^2 + K_{kl}K^{kl})\right]\right)  \nonumber \\
&&\qquad 
%
%+3s X \phi^i B_i ({\bar R} -s K^{[k}_kK^{l]}_l)
%
 %-6X C^{ij}\phi_j(K_{ik}{}^{;k} - K_{;i})
 %
  -C^{[i}_iC^j_jC^{k]}_k \phi_n
 %
 % \nonumber  \\\nonumber
%&&\qquad
 %
 +3s\phi_iB^i C^{[k}_kC^{l]}_l -6s  \phi_i C^i_j B^{[j}C^{k]}_k 
%
%+6XC \phi^i (K_{ij}{}{}^{;j}-K_{;i})
%
%\\\nonumber
%&&\qquad 
%
 -6s XC \phi_n K_{ij}K^{ij}
%
 %-6sX \phi^i B^j ({\bar R}_{ij} -s KK_{ij} +2s K_{ik}K^k_j)
 %
  -6XC^{ij} \phi^{k}K_{i[k;j]} 
  %
  %+6X B^i \phi^j K_{ik}K^k_j
  %
   \\\nonumber
&&\qquad -6sX\phi_n K_{ik}K^{kj}C^i_j   +3s(XK^{[i}_iK^{j]}_j\phi_n\phi_k)^{;k} +6 (X\phi_i)^{;i} \phi^{[j}_jK^{k]}_k\\\nonumber
&& \qquad +6X\phi^{[i}_i K^{j]}{}_{j;k}\phi^k + 3(Z \phi_n \phi_i)^{;i} {\bar R} -6(X\phi^i)^{[;j}K^{k]}{}_k\phi_{ij} -6s (X\phi_n)^{[;i}K^{j]}_j\phi_{ni}\\\nonumber
&&\qquad  +6 (X^2)^{[;i}K^{j]}{}_{j;i} -3(\phi_n \phi_i)^{;[i}\phi^j_j\phi^{k]}_k -6X_{;i}\phi_n{}^{;[i}\phi^{j]}_j -6X(\phi^iK_{ij})^{[;j}C_k^{k]}+6X B^{[i}K^{j]}{}_{j;i}\phi_n \\\nonumber
&&\qquad +6X K^{[i}{}_i{\bar R}_{jk}{}{}^{k]l}\phi^j \phi_l -6 \phi_n \phi^{[i}_i{\bar R}_{jk}{}{}^{k]l} \phi^j \phi_l - 6 Z \phi_n {\bar R}_{ij}\phi^{ij} - 12(Z\phi_n)_{;i} \phi_j {\bar R}^{ij} \\
&&\qquad -6s X{\bar R}_{ij}B^i \phi^j 
%
%+3(X^2)_{;i}(K^{ij}{}{}_{;j} - K^{;i}) 
%
+6 X\phi_{ni} B^{[i}K^{j]}_j \Bigg]
\ea
}
where we recall that $C_{ij}$ and $B_i$ are given by equations (\ref{Cij}) and (\ref{Bi}) respectively.
Note that all of these formulae  agree with those presented in \cite{gal-bterms}, save for the boundary curvature terms. It seems that the possibility of a non-trivial embedding and the resulting boundary curvature was not considered in \cite{gal-bterms}. It might be interesting to see what effect these additional terms have on the value of the on-shell Hamiltonian calculated in \cite{vish-ham}.

\section{Outlook} \label{sec:conc} 
Horndeski's general scalar-tensor theory \cite{horny} has received something of a renaissance in the last year.  It is no surprise that this has coincided with the development of galileons -- scalar Lagrangians with seemingly higher derivative interactions that preserve second order field equations.  These have a number of important applications ranging from consistent violations of the null energy condition \cite{null}, to soliton stabilisation \cite{solitons}. When coupled to gravity, such theories can exhibit self acceleration \cite{gal,bigal}, self tuning \cite{bigal}, and Vainshtein screening \cite{vainsh,gal,bigal,hirsute,def}.  These properties are inherited, of course, by Horndeski's generalisation, but Horndeski's theory can offer even more. It  also includes chameleons \cite{cham}, quintessence and k-essence \cite{kessence},  as well as accomodating  Higgs inflation \cite{higgs}. 

By computing the boundary terms and junction conditions for thin shells in Horndeski's theory, we have opened up the possibilty of further applications. The boundary terms, being the analogue of the Gibbons-Hawking term \cite{York,GH} in GR,  enable us to apply Euclidean path integral methods to the theory. Armed with the junction conditions one can in principle  construct Coleman-De Luccia instantons \cite{cd}, and use the well defined action to compute tunnelling rates.  Indeed, such  analyses may  capture salient features of tunnelling within the string landscape \cite{landscape}, at least  if  we can treat Horndeski's theory as a toy representation. 

The junction conditions will also enable us to study collapse of a spherical shell in a large class of modified gravity theories, along the lines initiated for the cubic covariant galileon in \cite{hirsute}, helping to develop our understanding of Vainshtein screening. Furthermore, given that our results do not depend on spacetime dimension, we are now in a position to study the dynamics of braneworlds  in a Horndeski bulk (for reviews of braneworld gravity, see \cite{roy, thesis}).  In particular it might be interesting to see what effect consistent violation of null energy \cite{null} in the bulk has on the dynamics of the brane, especially in view of \cite{opera}.

\acknowledgements{We would like to thank Paul Saffin for useful discussions, and for floating the idea of an analogous project when we first stumbled across Horndeski's theory. AP was funded by a Royal Society University Research Fellowship, and VS by a University of Nottingham  VC Scholarship for Research Excellence.}

%%%%%%%%%%%%%%%%%%%%%%%%%%%%%%%%%%%%%%%%%%%%%%%%%%%%%%%%%%%%%%%%%%%%

\appendix

\section{Notations and Identities used} \label{sec:useful}
This section contains  identities that were useful in deriving many of the formulae presented in this paper.  Recall that we are using bulk coordinates $x^a$, and boundary coordinates $\xi^i$, and  we may think of the boundary as an embedding $x^a=X^a(\xi)$. This defines tangent vectors $\del_i X^a$, each of which is orthogonal to the unit outward point normal $n^a$.\textbf{ Note that we have assumed that the normal vector $n^a$ is extended along geodesic such that, $ a^b \equiv n_a \nabla^a n_b =0$. This assumption does not affect the generality of the junction conditions since $a^b$ lies on the boundary and is continuous across the boundary.} The induced metric on the boundary is defined as
\be \nonumber
h_{ij}= \del_i X^a\del_j X^b g_{ab}|_{\del \cal M}
\ee
with the Lie derivative along the normal giving the extrinsic curvature 
\be \nonumber
K_{ij} =\frac{1}{2} {\cal L}_n h_{ij}
\ee
We have repeatedly made use of the following expressions
\ba
 B_i &=&  s\del_i X^a n^b  \nabla_a\nabla_b \phi= s \bar D_i \nabla_n\phi -s K_{ij}\bar D^j \phi \nonumber\\\nonumber
C_{ij} &=& \del_i X^a \del_j  X^b \nabla_a\nabla_b \phi = \bar D_i \bar D_j \phi +s K_{ij}  \nabla_n \phi
\ea
and the following identites
\ba
n^an^b  G_{ab} &=& -\frac{s}{2}\left[\bar R -s K^{[i}_iK^{j]}_j \right] \nonumber \label {gc1}\\ \nonumber
n^a\del_i X^b G_{ab} &=& \bar D^jK_{ij} - \bar D_i K \label{gc2} \\\nonumber
\del_i X^a \del_i X^bG_{ab}&=& \bar G_{ij} -s \left[ KK_{ij} - 2K_{ik}K^k_j - \frac{1}{2} h_{ij} (K^2 + K_{kl}K^{kl}) + {\cal L}_n K_{ij} - h_{ij}  {\cal L}_n K \right] \label{gc3} \\\nonumber
n^an^b R_{ab} &=& - {\cal L}_nK - K_{ij}K^{ij}\\\nonumber
n^an^c \del_i X^b \del_j X^d R_{abcd} & =& - {\cal L}_nK_{ij} + K_{ik}K^k_j\\\nonumber
\del_i X^a \del_i X^b R_{ab}&=& \bar R_{ij} -s KK_{ij} +2s  K_{ik}K^k_j -s {\cal L}_nK_{ij}\\\nonumber
R &=& \bar R -s K^2 -s K_{ij}K^{ij} -2s {\cal L}_nK
\ea

\end{document}